\documentclass[conference, 10pt]{IEEEtran}
\usepackage{caption}
\usepackage{subcaption}
\usepackage{graphicx}
\usepackage{color}
\usepackage{amsmath}
\usepackage{amssymb}
\usepackage{color}
\usepackage{epstopdf}
\usepackage{cite}
\usepackage{array}
\usepackage{multirow}
\usepackage{hhline}
\usepackage{epsfig}
\usepackage{latexsym}
\usepackage{mathrsfs}
\usepackage{float}
\usepackage[english]{babel}
\usepackage{color}
\usepackage[utf8]{inputenc}
\usepackage{pbox}
\usepackage{cleveref}
\usepackage[english]{babel}
\usepackage[T1]{fontenc}
\usepackage[utf8]{inputenc} 
\usepackage{booktabs}
\usepackage{lipsum}
\newcommand\blfootnote[1]{%
	\begingroup
	\renewcommand\thefootnote{}\footnote{#1}%
	\addtocounter{footnote}{-1}%
	\endgroup
}

\begin{document}
\title{MIMO Antenna Elements Effect on Chassis Modes}
\author{\IEEEauthorblockN{Asim Ghalib and Mohammad S. Sharawi}
	\IEEEauthorblockA{Electrical Engineering Department\\
		 King Fahd University of Petroleum and Minerals (KFUPM)\\
		email:\{\textit{g201405200,msharawi}\}@\textit{kfupm.edu.sa}\\}
}

\maketitle
\begin{abstract}
In this paper, a 4-element printed multiple-input-multiple-output (MIMO) loop antenna having a bandwidth of 400 MHz is proposed. The effect of the MIMO antenna on the chassis modes system is analyzed via the theory of characteristic modes (TCM), and it is shown that chassis CM are not enough for the full analysis. A defected-ground-structure (DGS) is also proposed to enhance the isolation between the antenna elements by blocking the coupling modes.
\end{abstract}
\IEEEpeerreviewmaketitle

\blfootnote{This work is supported by DSR-KFUPM under Project no. KAUST-002.}
\vspace{-1 em}
\section{Introduction}

Theory of characteristic modes (TCM) was initially developed by Garbacz \cite{Garbacez_thesis} but it gained importance after it was revisited in \cite{CabedoFabres2007}. Characteristic modes (CM) are given \cite{Ghalib2018a,Ghalib2018}
\begin{equation}
[X]J_n = \lambda_n [R]J_n
\end{equation}
where, $X$ and $R$ represents the imaginary and real part of the impedance matrix $Z$. $\lambda_n $ is the eigenvalue corresponding to eigenfunction i.e. current density ($J_n$). TCM is widely used for the design and analysis of various type of antennas and designs made with TCM are more robust \cite{Vasilev2016c,Ghalib2017d,Ikram2016c,Ghalib2017e,Ghalib2016a}. 


One of the performance metrics of multiple-input-multiple-output (MIMO) is port isolation \cite{Sharawi2013}. Alot of empirical methods such as decoupling networks, parasitic element, neutralization technique and defected ground structure (DGS) are proposed in literature to enhance the isolation. All the method relies on empirical approach. TCM was also used to improve the isolation between MIMO antennas \cite{Li2012,Ghalib2017,Li2012a}. In \cite{Li2012}, it was observed that if we restrict our study to less than 1 GHz, only one chassis mode will be present. So, if two antennas are placed in a such way that one of them excite the chassis while the other antenna does not excite the chassis, this can yield better isolation. So, one antenna was placed at the electric field maxima (coupling to chassis) and the other one at electric field minima (not coupling to chassis), as a result 5dB isolation improvement was achieved. 

Several LTE bands use frequencies greater than 1.5 GHz, thus it means that we have to deal with more than one chassis mode. Let us assume the case when we have two chassis modes (for a frequency greater than 1.83 GHz and a chassis size of 120x60 mm$^2$), we can see that there are no locations on the chassis where the modes have electric field minima or maxima at the same time. Normally when one mode has maxima the other mode has minima at that location. This means that the method proposed in \cite{Li2012} has limitations. Secondly, only the chassis modes are discussed  and analysis are made based on it. The effect of the single antenna and MIMO antennas on the chassis modes are totally ignored. In this work, we try to answer these effects. Besides we present a 4-element MIMO loop antenna with enhanced isolation.
\section{Analysis Process}
\begin{figure}[h]
\begin{center}
	\noindent
	\includegraphics[width= 9.0cm, height= 4.5 cm]{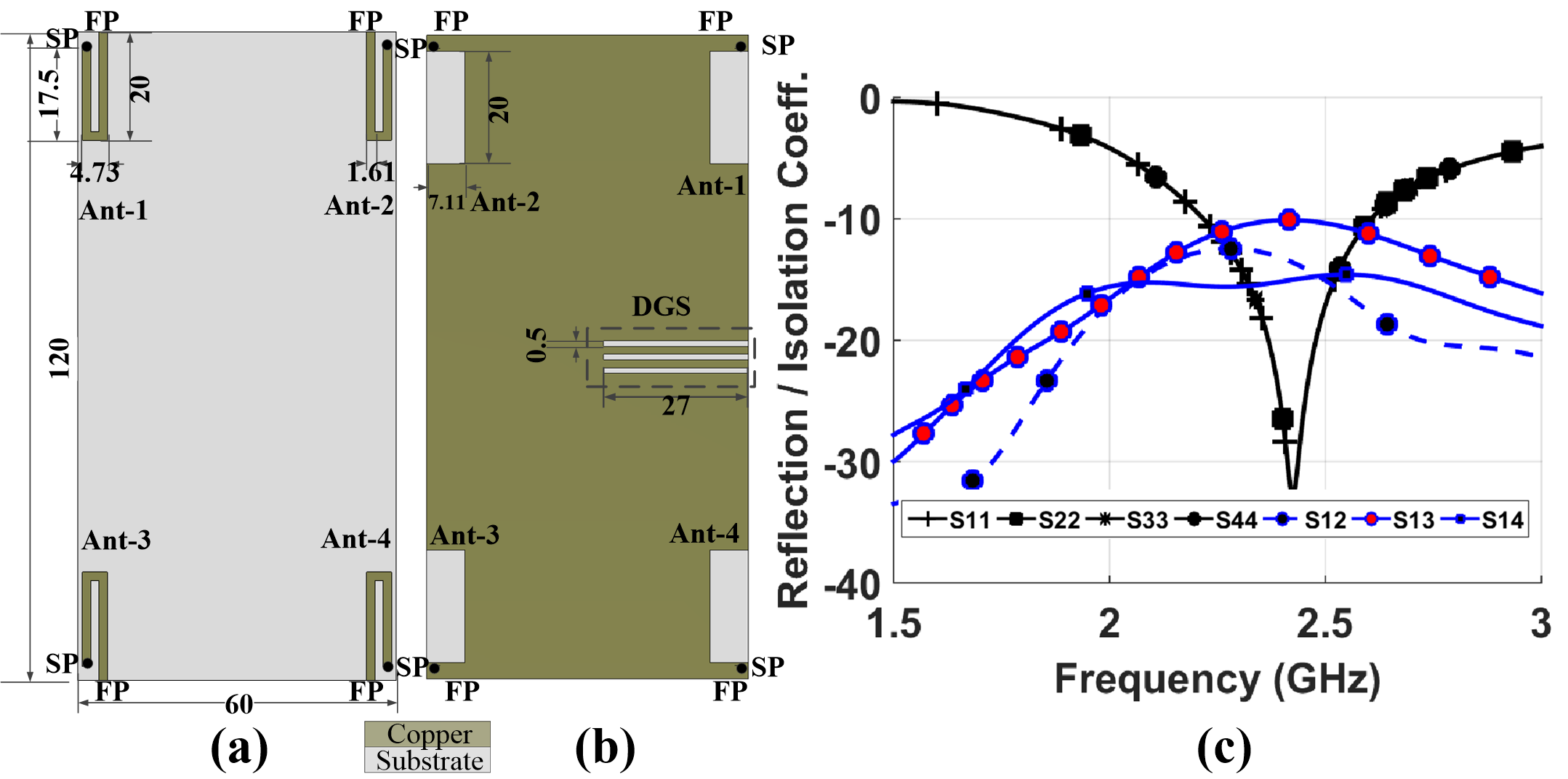}
	\caption{4 element MIMO loop antenna, where (a) top view, (b) bottom view and (c) Reflection and isolation curves without the DGS. SP and FP refers to the shorting and feeding point respectively. All dimensions are in mm.}
	\label{Geometry}
\end{center}
\end{figure}

To analyze the effect of multiple antennas, we followed a systematic procedure. We considered a normal mobile chassis of 120x60 mm$^2$ dimensions. The effect of 1-element, 2 and 4-element loop antennas on the chassis modes was investigated. For brevity reasons, the 1-element and 2-element printed loop antenna cases are not shown separately. The top view and bottom view of the proposed design is shown in the Fig. \ref{Geometry}(a) and \ref{Geometry}(b), respectively, while the reflection and isolation curves are shown in the Fig. \ref{Geometry}(c). The antenna is designed on an FR4 substrate having a dielectric constant of 4.0 and substrate thickness of 0.8 mm.

\begin{figure}[h]
	\vspace{-1.3 em}
	\begin{center}
		\noindent
		\includegraphics[width= 8.5cm, height= 4.5 cm]{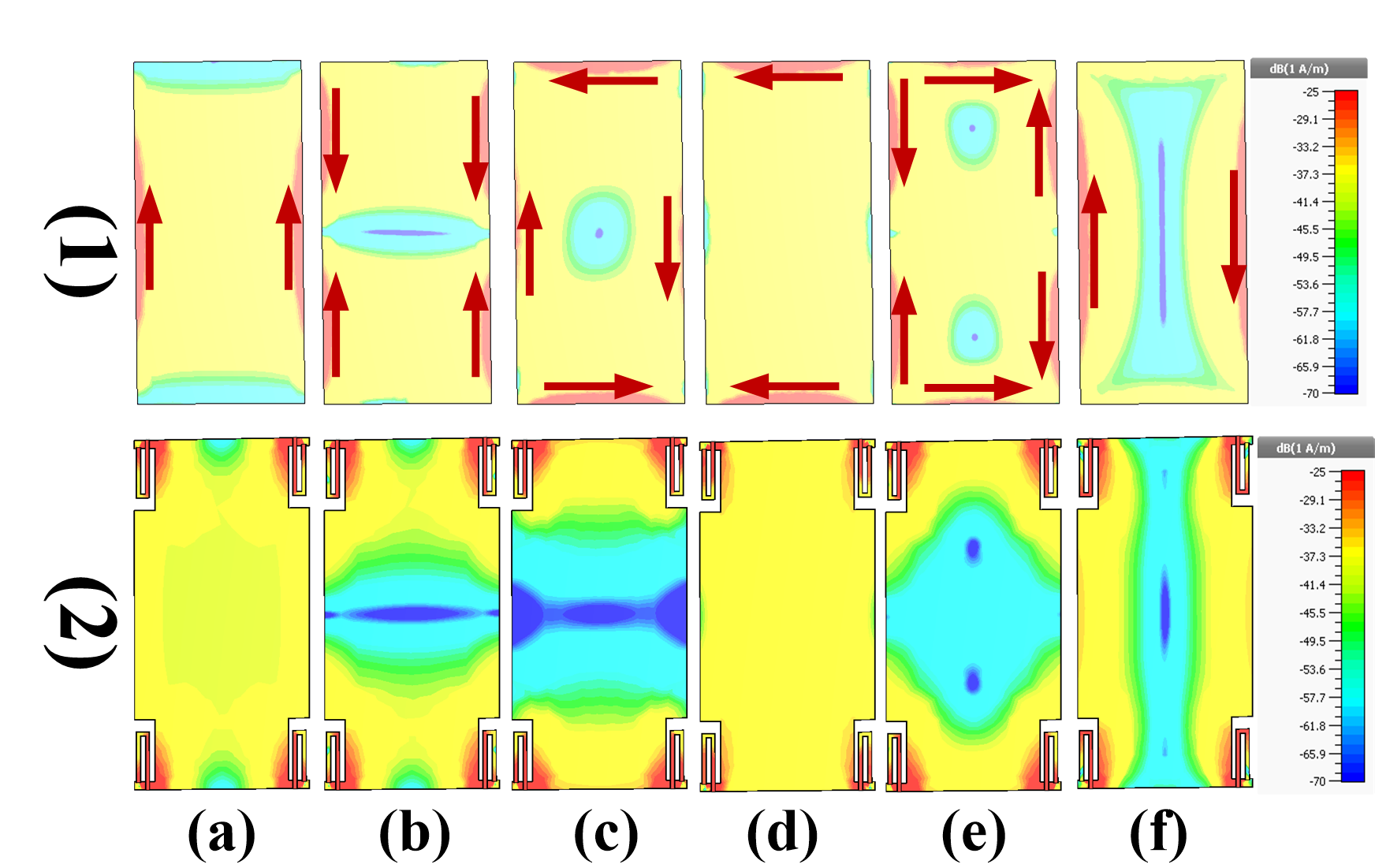}
		\caption{CM current distribution, where (1) ground plane, (2) 4-element MIMO loop antenna.(a)-(f) represents mode 1 to 6.}
		\label{CM_current_distr}
	\end{center}
\vspace{-2.75 em}
\end{figure}

\begin{figure}[h]
	\begin{center}
		\noindent
		\includegraphics[width= 8.0cm, height= 5.5 cm]{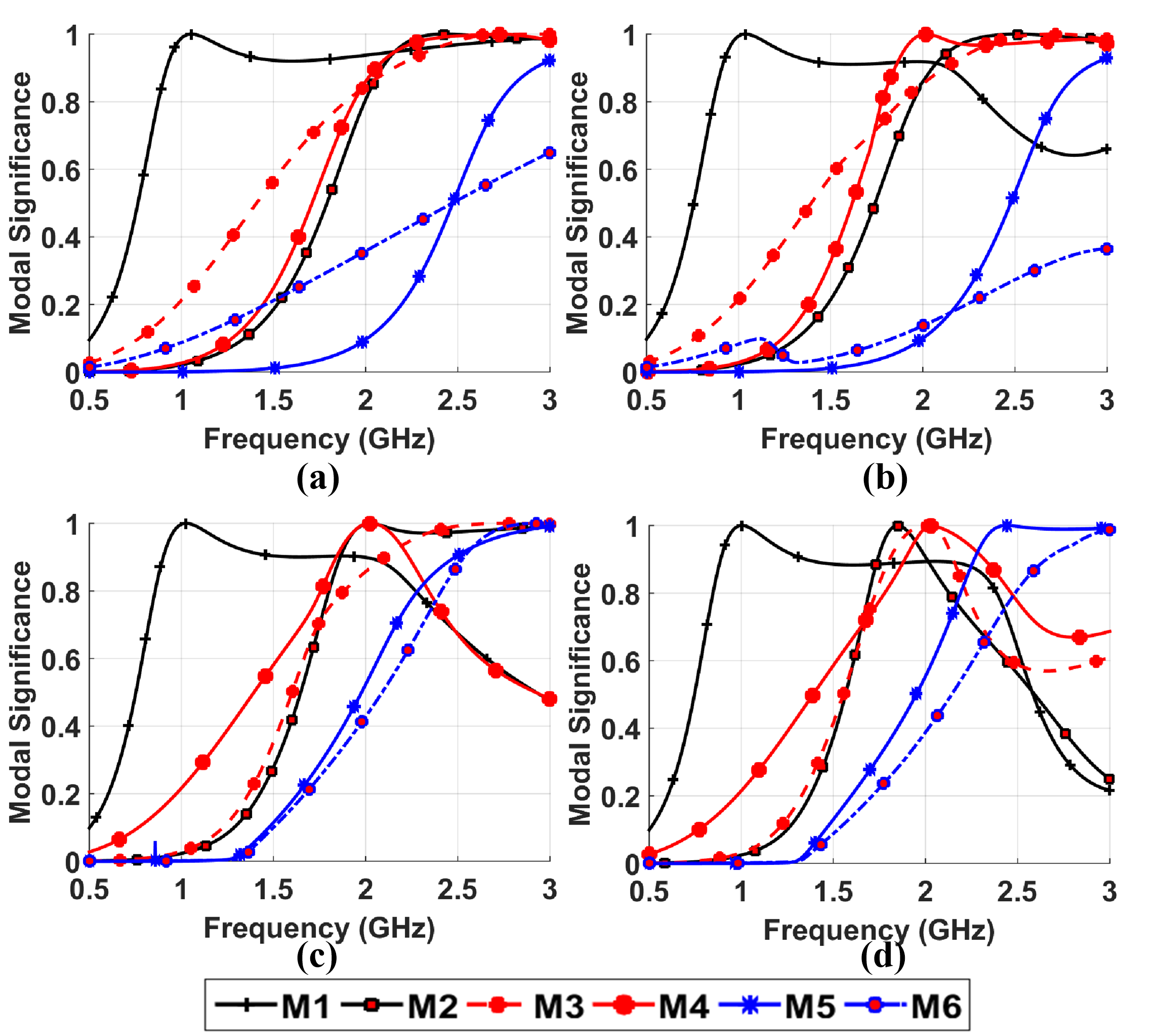}
		\caption{MS curve where, (a) chassis (120x60 mm$^2$), (b) 1 element printed loop antenna on a chassis, (c) 2-element printed MIMO loop antenna installed on shorter edge of the chassis and (d) 4-element printed MIMO loop antenna.}
		\label{MS_curve}
	\end{center}
\end{figure}

The CM and the modal significance (MS) curve of the ground plan are shown in the Figs. \ref{CM_current_distr}(1a-1f) and \ref{MS_curve}(a) respectively. The MS curves of the chassis modes are significantly affected after the introduction of the antenna as shown in the Fig. \ref{MS_curve}(b). For brevity the current distributions are not shown. Mode 1 radiating BW is severely affected. The other modes i.e. mode 2 to 6 are slightly affected. This proves that the presence of the antenna affects the CM. In the presence of the antenna all the current maxima shift to the antenna element. 

The MS curves of a 2-element MIMO antenna placed at the shorter edge of the chassis are shown in the Figure \ref{MS_curve}(c). We can observe that as compared to the case of chassis and single antenna, the CM of the chassis are affected more. We can observe that the MS curves of the modes 1, 4 and 6 are severely affected. Mode 4 seems to be effected because the antennas are placed at the corners of the shorter edge and mode 4 has current maxima at the shorter edge of the chassis. Mode 6 has started contributing to the radiating BW. The amount of affect on mode 1 seems to be same for the single and 2-element cases. 

The CM and the MS curves for the 4-element MIMO case are shown in Fig. \ref{CM_current_distr}(2a-2f) and \ref{MS_curve}(d) respectively. As expected the 4-element MIMO case has highly affected the chassis modes as compared to 1 and 2-element MIMO cases. The current maxima lies on the antenna element and the radiating BW of almost all the modes are affected.This validates that the affect of the antennas especially multiple antennas cannot be ignored and we need to take them into account while analyzing them. 

\begin{figure}[h]
	\begin{center}
		\noindent
		\includegraphics[width= 9.0cm, height= 4.5 cm]{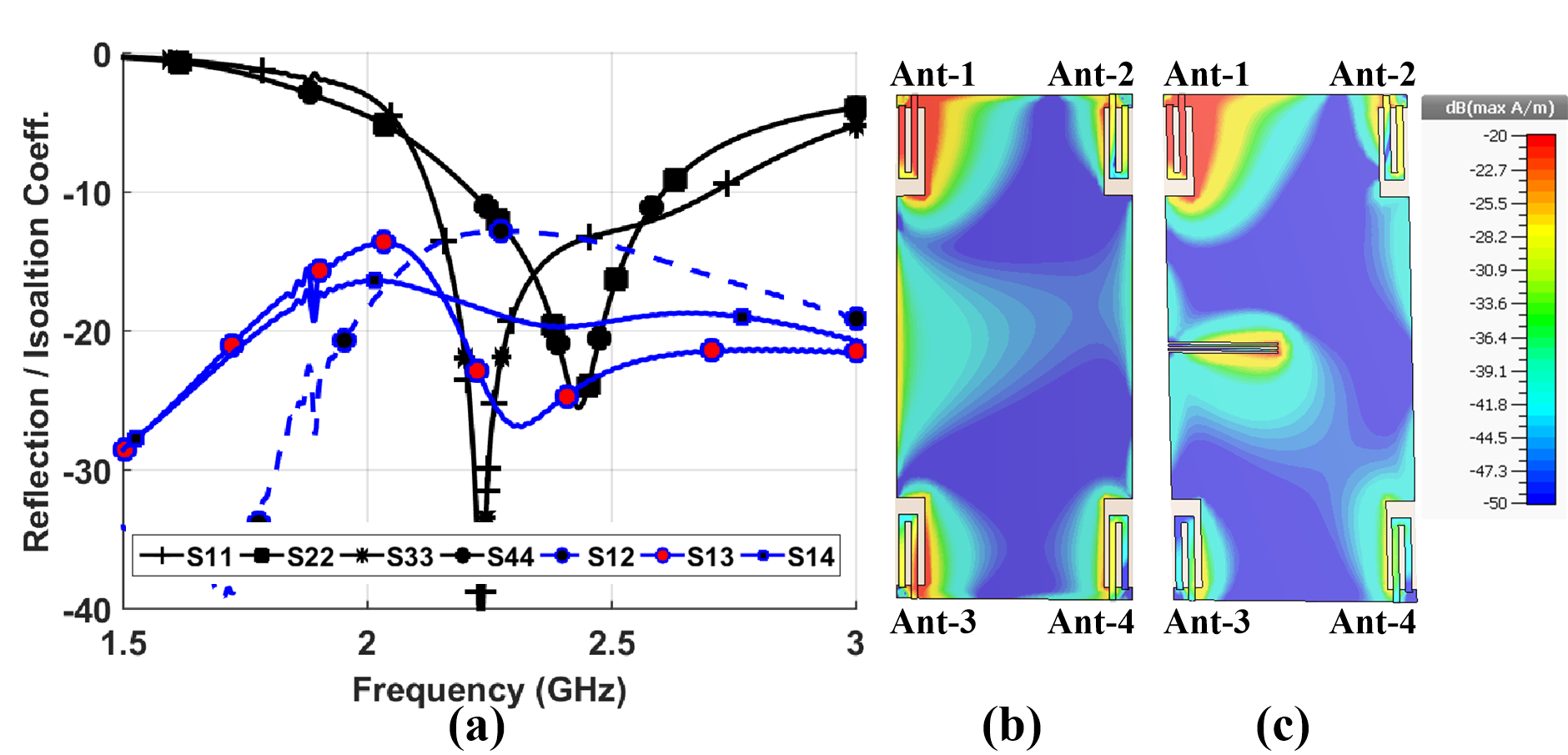}
		\caption{4-element MIMO loop antenna, where (a) Reflection and isolation curves, (b) current distribution when antenna 1 is excited in the absence of the DGS and (c)  presence of DGS.}
		\label{Isolation}
	\end{center}
\end{figure}

An isolation of 10 dB (poor) can be observed between Ant-1 and 3 as shown in the Fig. \ref{Geometry}(c). If we carefully observe the CM current distribution of the chassis and 4-element MIMO in Figs. \ref{CM_current_distr}(1a-2f), we can observe that modes 1, 5 and 6 are contributing to the coupling while modes 2 and 3 are not contributing because they have a current null (observe the blue spot in the middle of the chassis). This means to improve the isolation we need to block the coupling modes while the non-coupling modes will remain unaffected. Remember that in the impedance BW of interest, only the first four modes are present in the radiating BW and all the modes have current maxima across the antenna element. So, placing a defected ground structure (DGS) at the middle of the chassis will not affect the non-coupling mode but will block the coupling mode. The proposed DGS is shown in Fig. \ref{Geometry}(b). It enhanced the isolation by 11 dB i.e. from -11 dB to -22 dB as shown in the Fig. \ref{Isolation}(a). It slightly shifted the resonance of Ant-1 and 3 but the effective BW is still from 2.2 GHz to 2.6 GHz i.e. 400 MHz on a VSWR<2 criteria. We can observe in Fig. \ref{Isolation}(b) and \ref{Isolation}(c) that the current coupling has been significantly reduced.

\section{Conclusion}
The effect of MIMO antennas on mobile chassis cannot be ignored because they severely affect the chassis modes. In the presence of the antenna the current distribution shifts to the antenna elements. A DGS was proposed to stop the coupling modes that enhanced the isolation of a 4-element printed loop based MIMO design by 11 dB.
\bibliographystyle{IEEEtran}

\bibliography{IEEEabrv,library}

\end{document}